\documentclass[pra,twocolumn,amsmath,amssymb]{revtex4-1}

\usepackage{graphicx}
\usepackage{dcolumn}
\usepackage{bm}
\usepackage{comment}

\begin{document}

\title{Traveling pulses in Class-I excitable media} 

\author{Andreu Arinyo-i-Prats$^{1,2}$}
\author{Pablo Moreno-Spiegelberg$^1$}
\author{Manuel A. Mat\'ias$^1$}
\author{Dami\`a Gomila$^1$}
\email{damia@ifisc.uib-csic.es}
\affiliation{$^1$IFISC (CSIC-UIB), Instituto de F\'{\i}sica Interdisciplinar y Sistemas Complejos,  E-07122 Palma de Mallorca,  Spain \\
$^2$ Institute of Computer Science, Czech Academy of Sciences, 182 07 Praha 8, Czech Republic}
\date{\today}

\begin{abstract}
We study Class-I excitable $1$-dimensional media showing the appearance of propagating traveling pulses. We consider a general model exhibiting Class-I excitability mediated by two different scenarios: a homoclinic (saddle-loop) and a SNIC (Saddle-Node on the Invariant Circle) bifurcations. The distinct properties of Class-I with respect to Class-II excitability infer unique properties to traveling pulses in Class-I excitable media. We show how the pulse shape inherit the infinite period of the homoclinic and SNIC bifurcations at threshold, exhibiting scaling behaviors in the spatial thickness of the pulses that are equivalent to the scaling behaviors of characteristic times in the temporal case.
\end{abstract}
\maketitle

Excitability is a nonlinear dynamical regime that is ubiquitous in Nature. Excitable systems, having a stationary dynamics, are characterized by their response to external stimuli with respect to a threshold. Thus, stimuli below the threshold exhibit linear damping in their return to the fixed point, while stimuli exceeding the threshold are characterized by a nontrivial trajectory in phase space before returning to the fixed point. This different response to external perturbations confers excitable systems with unique information processing capabilities, and also the possibility of filtering noise below the threshold level. Moreover, excitable media, that are spatially extended systems that locally exhibit excitable dynamics, can propagate information, as happens in neuronal fibers \cite{KeenerSneyd98} or the heart tissue \cite{zykov90}.

From a dynamical systems point of view, excitability is typically associated to the sudden creation (or destruction) of a limit cycle, whose remnant traces in phase space constitute the excitable excursion \citep{IzhikevichDSN}. 
The route (i.e. bifurcation) through which a limit cycle is created or destroyed, leads to differences in the excitable dynamics. A basic classification of excitability is in two classes (types), depending on the response to external perturbations \citep{IzhikevichDSN}. Class-I excitable systems are characterized by a frequency response that starts from zero, leading to a (theoretically) infinite response time at the threshold. On the other hand, Class-II excitable systems are characterized by a frequency response that occurs in a relatively narrow interval, and thus the response time is bounded. Regarding the bifurcations that originate these two types of excitable systems, Class-I excitability occurs in certain bifurcations that involve a saddle fixed point when creating/destroying a limit cycle, as are the cases of a homoclinic (saddle-loop) or SNIC (Saddle Node on the Invariant Circle), also known as SNIPER (Saddle-Node Infinite Period), bifurcations. In turn, Class-II excitability is mediated by transitions involving a Hopf bifurcation such that in relatively narrow parameter range a large amplitude cycle is created, as is the case of subcritical Hopf bifurcations (typically close to the transition from sub- to supercritical Hopf bifurcation) and also the case of a supercritical Hopf bifurcation followed by a canard, i.e, a sudden growth of the cycle happening sometimes in fast-slow systems. This excludes the regular supercritical Hopf bifurcation, characterized by a gentle growth of the limit cycle amplitude.

Excitable media, obtained by coupling spatially dynamical systems that are locally excitable, show different regimes in which local excitations exceeding a threshold can propagate across the medium \cite{Mikhailovbook,Meron92}. 
Many studies have been carried out in Class-II excitable media, but much less is know about pulse propagation in the Class-I case. Excitable regimes are relevant in a number of physical, chemical and biological systems, namely semiconductor lasers, chemical clocks, the heart, and signal transmission in neural fibers. $1$-D pulse propagation is also behind nerve impulse transmission. 

In one spatial dimension both pulse propagation and periodic wave train regimes are found in Class-II \citep{KeenerSneyd98,Mikhailovbook}, and their instabilities have been characterized for representative models \cite{Zimmermann1997,OrGuil2001}. In $2$ and $3$ spatial dimensions further regimes are reported in Class-II, like spiral waves, including spiral breakup leading to spatiotemporal chaos \citep{Bar1999} and Winfree turbulence \citep{alonso2003taming}. These transitions are relevant in the study of excitable waves in heart tissue \cite{zykov90,panfilov98,Bar2019}, where they are associated to certain patologies.

Class-I excitability is much less studied and appears in models of population \cite{Baurmann2007,Iuorio2020} or neural \cite{Montbrio19} dynamics, and evidence of pulse propagation has been found in seagrasses \cite{ruiz2019}. Class-I pulse propagation has also recently been studied in arrays of coupled semiconductor lasers \cite{Alfaro2020}. The different properties of Class-I and Class-II excitability, specially the divergence of the period at threshold, can significantly modify the properties of spatiotemporal structures in excitable media. In this Letter we characterize traveling pulses in Class-I excitable media and show their distinct properties and instabilities. 

To address this problem we propose a general model based on
the normal form of a codimension-$3$  bifurcation \cite{dumortier2006bifurcations} which 
is the simplest continuous model one can write with Class-I excitable behavior that can be accessed either through an homoclinic (saddle-loop) or a SNIC bifurcation. To this normal form we add $1$-D diffusion to study spatial propagation:
\begin{align}
\label{Spsist}
\partial_t u &=v + D_{11} \partial_{xx} u  \\
\partial_t v &= \varepsilon_1 u^{3}+\mu _{2} u+\mu _{1} +v(\nu +bu- \varepsilon_2 u^{2}) +D_{22} \partial_{xx} v\ . \nonumber
\end{align}
We choose $\varepsilon _{1}=\varepsilon _{2}=-1$ to assure asymptotic stability,
and in the present work we fix the parameters $\nu = 1 $, $b = 2.4$ and $D_{11}=D_{22}=1$, considering $\mu_{1}$ and $\mu_{2}$ as control parameters. 

 \begin{figure}
\includegraphics[width=0.5\textwidth]{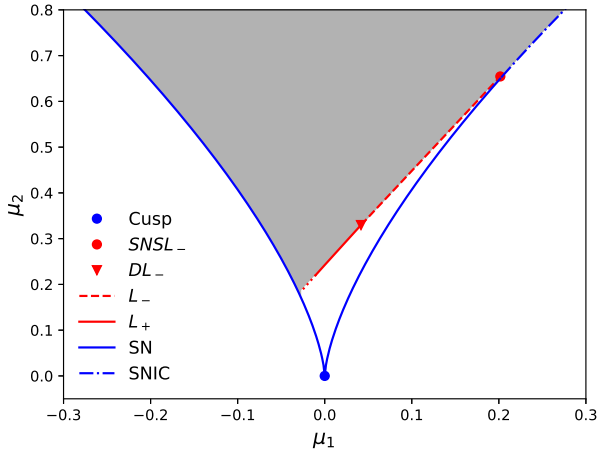}
\caption{Simplified phase diagram of the temporal system ($\partial_{xx}u=\partial_{xx}v=0$) in the parameter space ($\mu_1,\mu_2$). The diagram is organized by three main codimension-$2$ points: i) a Cusp (blue dot), where the Saddle-Node bifurcation lines (blue) meet, ii) a
Saddle-Node Separatrix Loop bifurcation (SNSL$_-$, red dot) on which a SNIC (blue dot-dashed line) and iii) a homoclinic of a stable cycle ($L_-$, red dashed line) bifurcation lines join, and ii) a bifurcation (DL$_-$, red triangle) in which $L_-$ meets with a homoclinic bifurcation of an unstable cycle ($L_+$, red solid line). The dotted end of $L_+$ indicates that the line continues (not shown) but it is not relevant for this work.}
\label{fig:saddle}
\end{figure}

The dynamical regimes exhibited by the temporal system (local dynamics), which is the system that describes the time evolution of homogeneous solutions  ($\partial_{xx} u=\partial_{xx} v=0$), are shown in Fig. \ref{fig:saddle}, where only the most relevant transitions to our study are shown.
The fixed points of this temporal system have $v^{\star}=0$ and $u^{\star}$ being determined by the solutions to the cubic equation formed by the first $3$ terms of the second equation in (\ref{Spsist}), that corresponds to the normal form of the cusp codimension-$2$ bifurcation.
The two blue lines are saddle node bifurcations that mark the boundary between the inside region with $3$ real roots
and the outside region with $1$ real plus a pair of complex conjugate roots, these $2$ lines joining at a cusp point in $\mu_1=\mu_2=0$ (blue circle in Fig.~\ref{fig:saddle}) with a triple degenerate root. 

There are other codimension-$2$ points, in addition to the cusp, that organize the scenario of interest to our study. One of them is the SNSL$^-$ point (Saddle-Node Separatrix Loop) \cite{Schecter1987}, (red circle in Fig.~\ref{fig:saddle}), that is characterized by a nascent (i.e. with a zero eigenvalue) homoclinic (saddle loop) bifurcation.
It is precisely from this SNSL$^-$ point that come up the two principal boundaries of the Class-I excitability region: a SNIC line (blue dot-dashed line), emerging upwards, and a homoclinic (saddle loop) line (red dashed line), $L_-$ downwards.

\begin{figure}[!ht]
\includegraphics[width=0.5\textwidth]{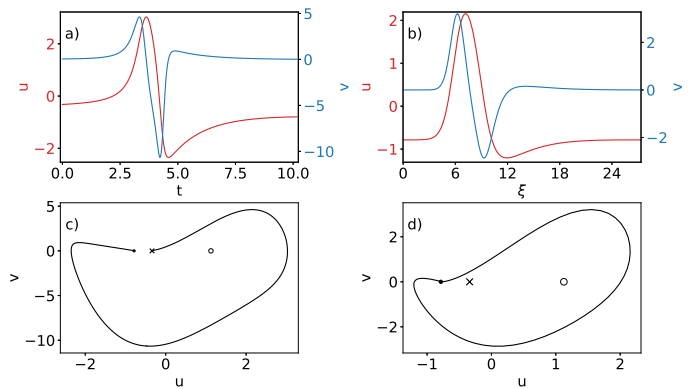}
\caption{Comparison of the temporal dynamics of an excitable excursion and the spatial dynamics of a $1$-D pulse sustained by this excitable dynamics. a) temporal excitable trajectory 
of $u$ and $v$ (spatially homogeneous system) starting from an initial condition just above the saddle point; c) representation of the same excitable excursion on the $(u,v)$ phase space;
b) stable $1$-D pulse as a function of the spatial coordinate in the moving reference frame; d) the same pulse in the $(u,v)$ (sub)phase space. Dot, cross, and circle in c) and d) indicate lower, middle and upper fixed points respectively. Here  $\mu_1 = 0.3$ and $\mu_2 =1.0$.}
\label{fig:fiducial_time}
\end{figure}

Another relevant codimension-$2$ point is the DL$^-$ (red triangle upside down), representing a homoclinic (saddle-loop) bifurcation to a neutral (resonant) saddle \footnote{Also known as resonant side switching point \cite{Champneys1994}, characterized by the fact that the absolute value of the leading eigenvalues of the saddle are equal.}, implying a transition between 
$L_-$ (that involves a stable cycle) and $L_+$ (red line, that involves an unstable cycle), and that
leads also to the emergence of a fold of cycles bifurcation line, not shown in Fig.~\ref{fig:saddle} for simplicity. The left SN line, both saddle-loop bifurcations ($L_-$ and $L_+$) and the SNIC curve delimit the Class-I excitable region (shaded grey area in Fig.~\ref{fig:saddle}). In this region a perturbation around the lower fixed point (black dot in Fig.~\ref{fig:fiducial_time}c) that crosses the threshold, i.e. the stable manifold of the middle fixed point (cross in Fig.~\ref{fig:fiducial_time}c), will trigger an excitable trajectory that comes back to the lower fixed point [panels a) and c) in  Fig.~\ref{fig:fiducial_time}].

The two above mentioned bifurcation lines, SNIC and $L_-$ 
are the ones of special interest to our study, as they mediate two different ways of entering the Class-I excitable region, cf. \citep{Gaspard1990,Jacobo2008}, and it will be reflected in the behavior of the pulses to be considered below. Furthermore, there are other codimension-$2$ points not relevant for our analysis \cite{Pablounpub}.

By initializing the system with a strong enough localized perturbation around the lower homogeneous solution a pair of solitary (or traveling) pulses that propagate with fixed shape and constant and opposite velocities are generated. One of such pulses, the one moving to the left, is shown in Fig.~\ref{fig:fiducial_time}b). A convenient way of characterizing this pulse is using a moving reference frame, $\xi=x-ct$, where $c$ is the velocity of the pulse yet to be determined. In this coordinate system the partial differential equations (\ref{Spsist}) become ordinary differential equations, and in our case we get,
\begin{eqnarray}
\label{Spsistrav}
du/d\xi &=&u_{\xi}\qquad ;\qquad dv/d\xi =v_{\xi} \\
du_{\xi}/d\xi&=&-(v + c\, u_{\xi})  \nonumber \\
dv_{\xi}/d\xi&=&u^{3}-\mu _{2} u-\mu _{1} -v (1 +bu+ u^{2})-c\, v_{\xi}  \nonumber\ .
\end{eqnarray}
Trajectories of this system describe stationary solutions of (\ref{Spsist}) in the reference frame moving with velocity $c$ \cite{Jaibi2020}. Only bounded trajectories have a physical meaning. In particular, excitable pulses are represented in this system as homoclinic trajectories originating from the lower fixed point (panels b and d in Fig.~\ref{fig:fiducial_time}). $c$ is computed numerically simultaneously with the field profiles, and it varies weakly with parameters in the excitable region.

 \begin{figure}[!ht]
 \centering
\includegraphics[width=0.5\textwidth]{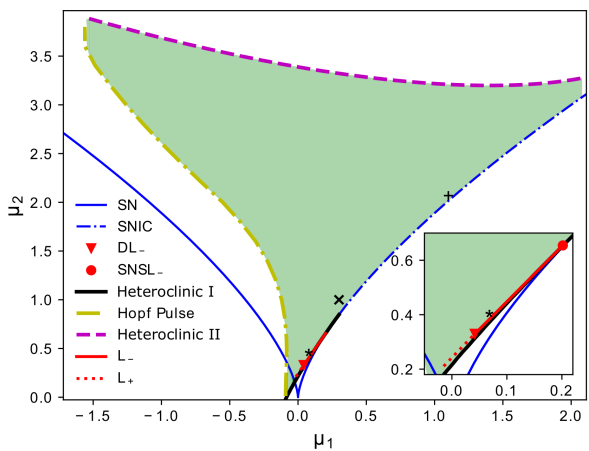}
\caption{\label{fig:stav_region} Phase diagram showing the bifurcation lines of $1$-D pulses. Traveling pulses are stable 
in the green region. The main instabilities discussed in this work are the Heteroclinic I (black line) and the SNIC (blue dot-dashed line). The other lines that bound the stability region are the Heteroclinic II and the Hopf of pulses (not discussed in this work). The SN, $L_-$ and $L_+$ lines are shown in order to compare the diagram with respect to Fig.~\ref{fig:saddle}. The $\times$, $*$ and $+$ symbols mark the parameter values studied in Figs.~\ref{fig:fiducial_time}, \ref{fig:2x2_Down_Heteroclinic}, and \ref{fig:2x2SNIC} respectively.}
\end{figure}

Although the dynamical system describing temporal dynamics for homogeneous solutions and the spatial dynamical system (\ref{Spsistrav}) are different, one may observe important similarities in their solutions. Roughly speaking, a traveling pulse somehow transcribes the temporal dynamics in space, such that the spatial profile of the pulse resembles the excitable trajectory in time. Thus, Fig.~\ref{fig:fiducial_time}(a) shows a excitable (open) trajectory in the temporal dynamics, while Fig.~\ref{fig:fiducial_time}(b) shows a excitable pulse in the spatial dynamics (a homoclinic orbit), for the same parameter values.
In Figs.~\ref{fig:fiducial_time}c) and \ref{fig:fiducial_time}d) the trajectories from Fig.~\ref{fig:fiducial_time}a) and  \ref{fig:fiducial_time}b) are represented in the $(u,v)$ phase space respectively. The similarity of panels c) and d) of Fig.~\ref{fig:fiducial_time}  anticipates the results presented in this work.

Next we analyze how the infinite period bifurcations leading to Class-I excitability in the temporal system, namely the homoclinic and SNIC bifurcations, affect the shape of the traveling pulses. To do so we study the domain of stability of pulses in the $(\mu_1,\mu_2)$ parameter space, shown in 
Fig. \ref{fig:stav_region},
where the cusp and saddle-node lines of Fig.~\ref{fig:saddle} are also included 
in the diagram for comparison.
This domain is delimited by several bifurcations at which the pulse is destroyed or made unstable:
Heteroclinics I and II, SNIC, and Hopf of pulses.
Here we focus on the Heteroclinic I and SNIC bifurcations, which are connected to the  $L_-$ (homoclinic) and the SNIC bifurcations of the temporal system respectively. 

\begin{figure}[!ht]
\includegraphics[width=0.5\textwidth]{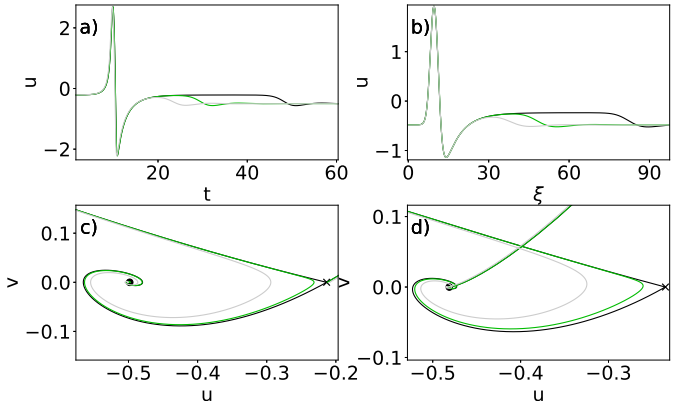}
\caption{\label{fig:2x2_Down_Heteroclinic} a) Divergence of the duration of the excitable excursion in the temporal system approaching the Homoclinic bifurcation, and b) divergence of the plateau in the pulses approaching the Heteroclinic I bifurcation. Here $\mu_2=0.4$ and $\mu_1=\mu_{1c}-\Delta\mu_1$ where the homoclinic bifurcation occurs at  $\mu_{1c} = 0.07560395587$ for the temporal system, and the Heteroclinic I at $\mu_{1c}=0.08107876002$ for the spatial systems, and
$\Delta\mu_1=10^{-3}$ (\textit{grey}), $\Delta\mu_1=10^{-5}$ (\textit{green}), $\Delta\mu_1=10^{-12}$ (\textit{black}) in all panels. 
Panels c) and d) show a zoom in of the most relevant region of the phase space $(u,v)$ for the temporal and spatial dynamics respectively.  }
\end{figure}

Let us first consider the Heteroclinic I curve, represented as a black line in Fig. \ref{fig:stav_region}.
Approaching this bifurcation
the pulse shape changes drastically, generating a plateau at the value of the middle (saddle) homogeneous solution (Fig.~\ref{fig:2x2_Down_Heteroclinic}b). 
As the spatial trajectory approaches the saddle point (through its stable manifold) there is a slowing down of the spatial dynamics, inherited from the temporal homoclinic (Fig.~\ref{fig:2x2_Down_Heteroclinic}a), that manifests as a plateau in the spatial profile. The plateau is more clear as one is very close to the bifurcation (black line).
Fig.~\ref{fig:2x2_Down_Heteroclinic}c) shows the temporal excitable excursion in the $(u,v)$ (sub)phase space, where it can be seen that the trajectory gets closer and closer to the saddle point, marked with a cross. The spatial counterpart (Fig.~\ref{fig:2x2_Down_Heteroclinic}d) behaves analogously, leading to the formation of a double heteroclinic at threshold, where the size of the plateau diverges. 

\begin{figure}[!ht]
\includegraphics[width=0.5\textwidth]{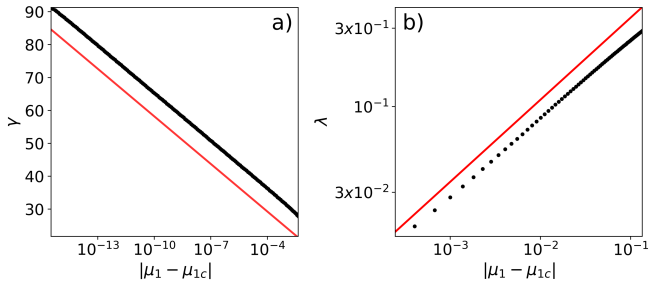}
\caption{\label{fig:Het_SNIC_tail} (a) Scaling of the pulse width, $\gamma$, approaching the Heteroclinic bifurcation. 
$\gamma$ is defined as the distance since the pulse separates $10^{-2}$ from the stable homogeneous solution until it comes back to the same distance. The expected scaling is $\gamma = \frac{1}{\lambda_1} log(|\mu-\mu_{1c}|)$ (shown in red in the plot), where $\lambda_1$ is the closest eigenvalue to zero of the middle fixed point in the spatial dynamics. 
b) Scaling of the eigenvalue, $\lambda$, that becomes $0$ at the SNIC as a function of the parameter distance to the SNIC bifurcation. The expected scaling is a power law with exponent $1/2$: $\lambda \propto \sqrt{|\mu_1-\mu_{1c}|}$ (red line).
}
\end{figure}

This slowing down has a characteristic logarithmic scaling law in the width of the plateau with respect to the parameter distance to the bifurcation \footnote{The logarithmic scaling law in the spatial coordinate $\xi$ is analogous to the temporal logarithmic scaling law of the homoclinic (saddle-loop) bifurcation \cite{Gaspard1990}.}, as shown in Fig.~\ref{fig:Het_SNIC_tail}a).
The red line represents the expected scaling slope from theory, that depends on the logarithmic parameter distance divided by the (independently obtained) leading unstable eigenvalue of the saddle point, and we can see that the agreement is perfect.

The fact that the Heteroclinic I bifurcation curve follows closely the $L_-$ and $L_+$ lines of the temporal dynamics (Fig.~\ref{fig:saddle}) and that the quantitative scaling for the width of the pulse has the same form of that of a homoclinic bifurcation in time indicate how the bifurcations of the temporal dynamics permeate the spatial dynamical description of the pulse, even though the connection is not straight forward from the equations.


The second instability of pulses that we  consider is the SNIC bifurcation (blue dash-dotted line in Fig.~\ref{fig:stav_region}). At this bifurcation a cycle is reconstructed when a saddle and a node collide, namely the lower and middle fixed points.
As the pulse approaches the SNIC, there is a slowing down in the approach to stable fixed point in the spatial dynamics, as the spatial eigenvalue tends to zero too.

\begin{figure}
\includegraphics[width=0.5\textwidth]{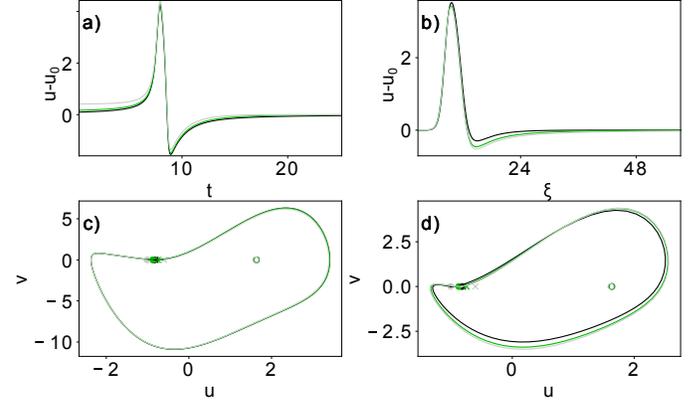}
\caption{\label{fig:2x2SNIC} Same as in Fig.~\ref{fig:2x2_Down_Heteroclinic} for the SNIC bifurcation. $u-u_0$ is plotted in panels a) and b), where $u_0$ is the bottom stable fixed point. Here $\mu_2 = 2.0$ and $\mu_{1c} = 1.0886621079036347$ with  $\Delta\mu_1 = -10^{-4},10^{-2},10^{-1}$ for \textit{black}, \textit{green} and \textit{grey} curves respectively. 
}
\end{figure}

In the temporal case the power law manifests in the divergence of the characteristic time to reach the stable fixed point
\cite{Jacobo2008}. Analogously, 
one would expect a power-law scaling of the pulse thickness, that would diverge at the onset of bifurcation.
However, due to the exponential approach to the saddle close to the bifurcation, the thickness of the pulse is not well defined. So we have turned to measure the approach rate, that is proportional to the leading eigenvalue, that becomes zero at the bifurcation.
This scaling is shown in Fig.~\ref{fig:2x2SNIC}b) and, as expected in a saddle-node bifurcation, it follows a power-law with exponent $1/2$. 
The behavior of the system at the other side of the bifurcation corresponds to a wave train, that in this case is the analog of a temporal periodic behavior, the SNIC marking, thus, a transition from wave trains to pulses.

In conclusion, we have shown the existence of $1$-D pulses in a model with excitable behavior corresponding to Class-I excitability mediated by two different bifurcations, SNIC and homoclinic (saddle-loop). 
We have characterized the region in parameter space in which the pulse is stable and two specific instability regimes of pulses: a heteroclinic that occurs for parameter values close to the temporal homoclinic and the SNIC, both defining Class-I excitability.
These instabilities exhibit at the transition the same scaling behaviors for the width of the pulses than those found in the period of the oscillations of the temporal case, namely logarithmic and power-law for the heteroclinic and SNIC bifurcations respectively, unveiling a profound relation between temporal systems and the spatiotemporal structures of partial differential equations.  Further instabilities of pulses in this system, marked as Heteroclinic II and a Hopf bifurcations in Fig.~\ref{fig:stav_region}, are beyond the scope of the present work and will be studied elsewhere \citep{Pablounpub}.

This work sets the ground to the study and explore spatio-temporal structures in Class-I excitable media, both $1$- and $2$-dimensional, a field unexplored hitherto.

We acknowledge financial support from FEDER/Ministerio de Ciencia, Innovación y Universidades - Agencia Estatal de Investigación through the SuMaEco project (RTI2018-095441-B-C22) and the María de Maeztu Program for Units of Excellence in R\&D (No. MDM-2017-0711). AAiP and PMS have contributed equally to this work.

%

\end{document}